\documentclass[aps, prl, twocolumn, notitlepage,superscriptaddress]{revtex4-2}
\usepackage{xspace}
\usepackage{graphicx}% Include figure files
\usepackage{dcolumn}% Align table columns on decimal point
\usepackage[usenames]{color}
\usepackage{slashed}
\usepackage[utf8]{inputenc}
\usepackage{amsfonts}
\usepackage{amsthm}
\usepackage{amsmath}
\usepackage{amssymb}
\usepackage{mathrsfs}
\usepackage{hyperref}
\usepackage{booktabs}
\usepackage{diagbox}
\usepackage{latexsym}
\usepackage{color}
\usepackage{bm}
\usepackage{CJK}
\usepackage{braket}

\usepackage{float}
\makeatletter
\let\newfloat\newfloat@ltx
\makeatother
\usepackage{algorithm}
\usepackage{algorithmicx}
\usepackage{algpseudocode}

\usepackage{caption}
\usepackage{subcaption}

\newcommand{\printfnsymbol}[1]{%
  \textsuperscript{\@fnsymbol{#1}}%
}

\newcommand{\drvParam}[0]{\boldsymbol{\gamma}}
\newcommand{\mixParam}[0]{\boldsymbol{\beta}}
\newcommand{\drvParamComp}[0]{\gamma}
\newcommand{\mixParamComp}[0]{\beta}

\newcommand{\clauses}[0]{C}

\begin{document}
\begin{CJK*}{UTF8}{gbsn}
\title{Quantum Dropout: On and Over the Hardness of Quantum Approximate Optimization Algorithm}

\author{Zhenduo Wang(王朕铎)}
\affiliation{International Center for Quantum Materials, School of Physics, Peking University, Beijing 100871, China}
\author{Pei-Lin Zheng}
\affiliation{International Center for Quantum Materials, School of Physics, Peking University, Beijing 100871, China}
\author{Biao Wu(吴飙)}
\affiliation{International Center for Quantum Materials, School of Physics, Peking University, Beijing 100871, China}
\affiliation{Wilczek Quantum Center, School of Physics and Astronomy, Shanghai Jiao Tong University, Shanghai 200240, China}
\affiliation{Collaborative Innovation Center of Quantum Matter, Beijing 100871, China}

\author{Yi Zhang}
\email{frankzhangyi@gmail.com}
\affiliation{International Center for Quantum Materials, School of Physics, Peking University, Beijing 100871, China}
\affiliation{Collaborative Innovation Center of Quantum Matter, Beijing 100871, China}

\begin{abstract}
A combinatorial optimization problem becomes very difficult in situations where the energy landscape is rugged, and the global minimum locates
in a narrow region of the configuration space. When using the quantum approximate optimization algorithm (QAOA) to tackle these harder cases,
we find that difficulty mainly originates from the QAOA quantum circuit instead of the cost function. To alleviate the issue, we selectively dropout the clauses defining the quantum circuit while keeping the cost function intact. Due to the combinatorial nature of the optimization problems,
the dropout of clauses in the circuit does not affect the solution. Our numerical results confirm  improvements in QAOA's performance
with various types of quantum-dropout implementation.
\end{abstract}

\maketitle
\end{CJK*}
\emph{Introduction}\textemdash A class of important real-world combinatorial optimization problems
%, including NP-complete or NP-hard problems,
cost exponential resources to solve on a classical computer \cite{Complexity1975, NPcomplete1990, Complexity2005}.
As the size of the problem increases, a solution becomes so costly that it is virtually impossible for even the world's largest supercomputers.
Although quantum computers have been shown to hold enormous advantages over classical computers on some specific problems \cite{Shor1999, Harrow2017, Bravyi2018, Arute2019, Zhong2020}, an important open question is whether a quantum computer can provide advantages and improve our stance on these difficult optimization problems.

The quantum approximate optimization algorithm (QAOA) is a hybrid quantum-classical variational algorithm designed to tackle ground-state problems, especially discrete combinatorial
optimization problems \cite{QAOA2014Farhi, QAOA2020PRX, QAOA2020PRApplied, Willsch2020, PhysRevLett.124.090504, Hsieh2019, Pagano2020, Streif2020, Sack2021, Medvidovic2021, Javier2022, QAOANPHard2021, PRXQuantum.2.010309, Harrigan2021, Amaro_2022, PhysRevA.97.022304}. It has been shown quite effective in many problems. However, we note that these studies mainly focus on randomly generated problems \cite{QAOA2014Farhi, QAOA2020PRX, QAOA2020PRApplied, Willsch2020, PhysRevLett.124.090504}, which may potentially concentrate on simpler cases and fail to represent the problem's categorical difficulty. In this work we try to address these nontrivial and
difficult cases with a strategy that we call quantum dropout.

\begin{figure}
\includegraphics[width=0.98\linewidth]{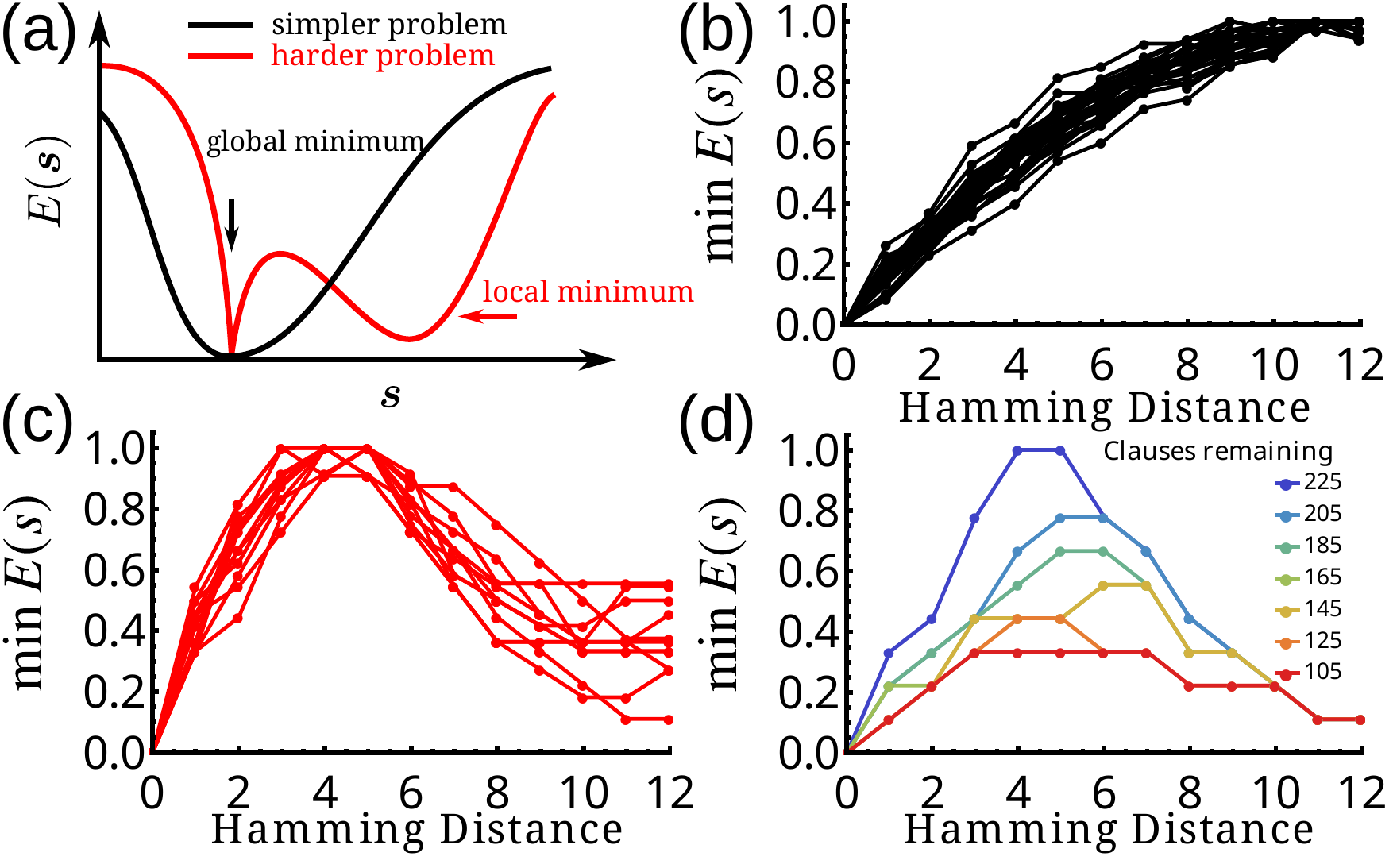}
\caption{(color online)(a) A schematic energy landscape. The global minimum is located in a large and smooth neighborhood for a simpler problem and a narrow region for a harder problem.
Normalized landscapes for NAE3SAT problems in terms of the hamming distance: (b) for simpler cases and (c) for harder cases. (d) The landscape of a harder problem becomes
smoother as more clauses are dropped out, improving the global minimum's standing.
Only the minimum of $E(\boldsymbol{s})$ for a specific hamming distance of $\bm{s}$ from the global minimum is shown for clarity. $n=24$.
}\label{fig:landscape}
\end{figure}

\begin{figure}
\includegraphics[width=0.98\linewidth]{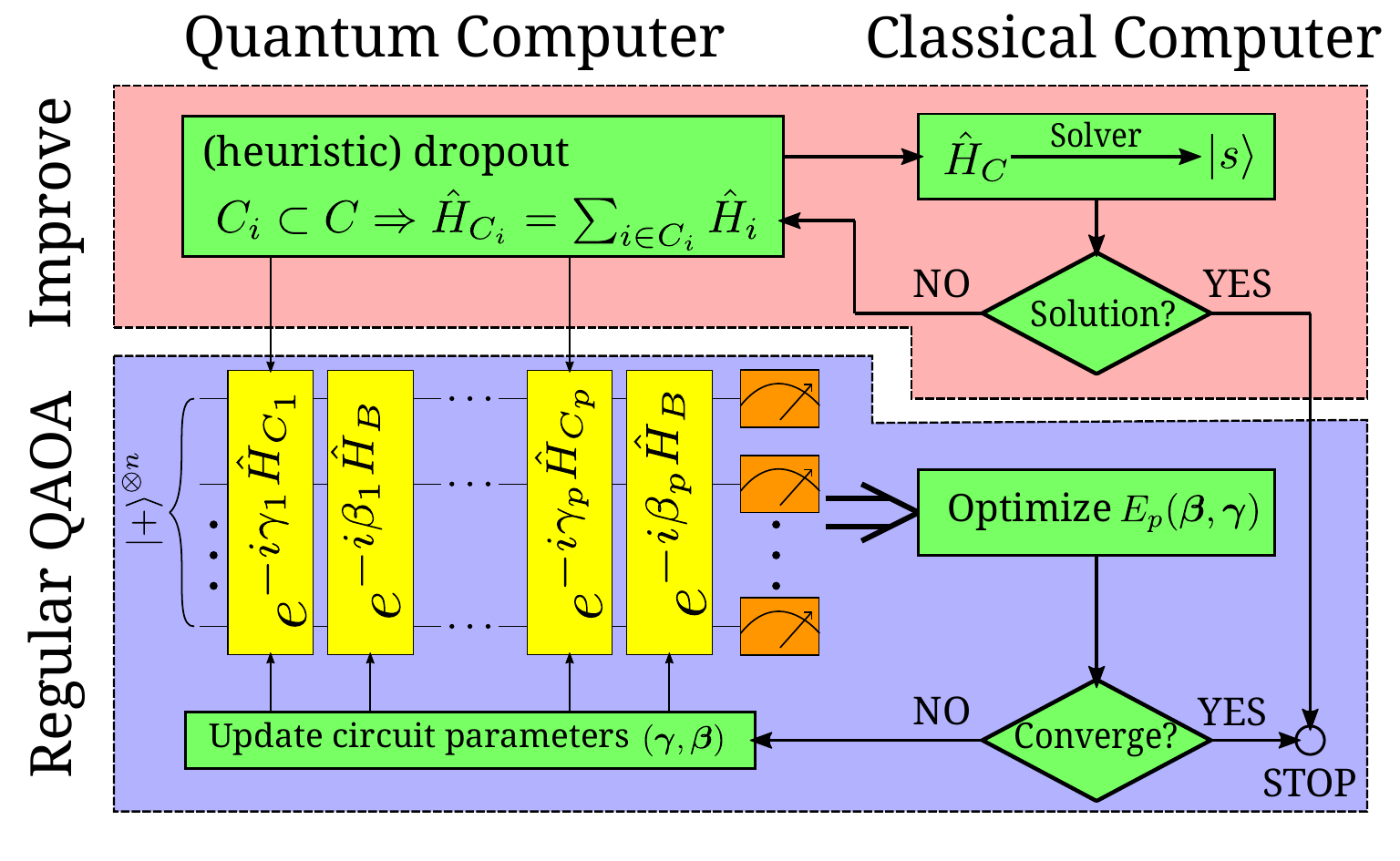}
\caption{(color online) QAOA is a hybrid optimization procedure that consists of a $2p$-layer quantum circuit evaluating a cost function and a classical algorithm optimizing the corresponding parameters $\mixParam, \drvParam$. Each bi-layer contains a driving layer ($e^{-i \hat H_C \drvParamComp_m}$) and a
mixing layer ($e^{-i \hat H_B \mixParamComp_m}$), where $m$ is the bi-layer index. In addition, we propose to check the target problem
with classical algorithms first to see whether QAOA is necessary and implement a dropout on the $\hat H_C$ clauses presented in each of the driving layers.}
\label{fig:qaoa}
\end{figure}

We focus on combinatorial optimization problems where the global minimum (ground state) $\boldsymbol{s}_{gs}$ has to satisfy each clause $\hat{H}_{i}$ in a given set $C$,
\begin{equation}
\hat H_C = \sum_{i \in C} \hat H_{i}\,.
\label{eq:Hcp}
\end{equation}
The difficulty of this problem depends on the energy landscape, the smoothness or roughness of) the objective function $\hat H_C(\boldsymbol{s})$ versus $\boldsymbol{s}$ with respect to its locality in the state space.
As illustrated in Fig. \ref{fig:landscape}a, when the global minimum locates in a large and smooth neighborhood, the problem is simpler
since its solution can be efficiently found with local-based searches such as simulated annealing (SA) \cite{SA1993} and the greedy algorithm \cite{Jungnickel1999}. When the minimum locates in a narrow region in a rugged landscape, the case is harder. This is further demonstrated
in Figs. \ref{fig:landscape}b and \ref{fig:landscape}c, which are the energy landscapes of simpler and harder cases
in the not-all-equal 3-SAT (NAE3SAT) problems \cite{Moret1988, Dinur2005}, respectively. Figs. \ref{fig:landscape}d shows that the energy landscape for harder cases becomes much smoother as more clauses are dropped out.
This observation, combined with our realization that a more rugged landscape makes the optimization of quantum circuit more costly,
leads us to a strategy, where we choose to dropout a portion of clauses from the quantum circuit, $\hat H_{C'} = \sum_{i \in C'\subset C} \hat{H}_{i}$, while keeping the original cost function to ensure the uniqueness of the global minimum. This strategy, as illustrated schematically in Fig. \ref{fig:qaoa}, utilizes the problems' combinatorial nature and has, in general, no trivial parallel in classical algorithms.
Our numerical results show that with quantum dropout QAOA can locate the ground states with an enhanced probability and little to no overhead even for harder cases,
paving the way towards practical QAOA for combinatorial optimizations.
In the deep artificial neural network, there is a vital technique called dropout, which keeps a random subset of neurons from optimization
for more independent neurons and thus suppresses over-fitting \cite{MLbook, Nitish2014}. Our quantum dropout echoes
this classical technique in spirit but is not a direct generalization.

\emph{QAOA and quantum dropout}\textemdash QAOA is to solve optimization problems such as Eq. \ref{eq:Hcp} by finding its ground state $|\bm{s}_{gs}\rangle$,
which is one of the $n$-qubit basis $\ket{\boldsymbol{s}}$ over the $\sigma^z$ configurations: $\sigma^z_r\ket{\boldsymbol{s}}= s_r\ket{\boldsymbol{s}}$. As shown in Fig. \ref{fig:qaoa},
QAOA offers a parameterized variational state:
\begin{equation}
    \ket{\mixParam, \drvParam} = e^{-i \hat H_B \mixParamComp_{p}} e^{-i \hat H_{C_p} \drvParamComp_{p}} \cdots e^{-i \hat H_B \mixParamComp_1} e^{-i \hat H_{C_1} \drvParamComp_1} \ket{+}^{\otimes n}, \label{eq:qaoa_appr}
\end{equation}
where $\hat H_B = -\sum_{r=1}^n \sigma^x_{r}$ with $\ket{+}^{\otimes n}$ being its ground state. In Ref. \cite{QAOA2014Farhi}, the original setup is inspired
by quantum adiabatic (annealing) algorithm \cite{Tadashi1998, Dorit2007, QA2008} such that $\hat H_{C_1} = \cdots \hat H_{C_p} = \hat H_C$. QAOA implements a quantum circuit
to efficiently evaluate the expectation value of Eq.~\ref{eq:qaoa_appr} as a cost function:
\begin{equation}
    E_p(\mixParam, \drvParam) = \braket{\mixParam, \drvParam | \hat H_C| \mixParam, \drvParam },  \label{eq:qaoa_cost}
\end{equation}
which is in turn optimized classically. The quantum circuit evaluates an exponential number of classical configurations simultaneously,
and with more layers the overlap $\langle \boldsymbol{s}_{gs} \ket{\mixParam, \drvParam}$ may become larger.
At the convergence,  $\ket{\mixParam, \drvParam}$ is measured in the basis of $\{\ket{\boldsymbol{s}}\}$.

In comparison with variational quantum eigensolver \cite{Peruzzo2014, Kandala2017, VQEreview2021}, QAOA possesses far fewer variational parameters - basically, the $2p$ variables of $\mixParam, \drvParam$. Sometimes, e.g., for simpler $\hat{H}_C$, QAOA may offer a sufficient approximation with merely a few layers $p\sim O(1)$; in general, however, we need $p \gtrsim O(n)$ for sufficient expressing power of $\ket{\mixParam, \drvParam}$ to encode $\boldsymbol{s}_{gs}$ \footnote{The scaling is partially due to the small Hamming distance of $\hat H_B$; however, such model locality is also the premise for arguments on quantum annealing, energy landscape, etc.}. However, a larger $p$ complicates the non-convex optimization of $E_p(\mixParam, \drvParam)$ \cite{QAOANPHard2021, Nonconvex2017, BP0, BP1, BP2},
especially for quantum circuits with harder $\hat{H}_{C}$, as we will see later. May we swap $\hat{H}_{C}$ for a simpler one in the quantum circuit? Unfortunately,
the answer is negative - a quantum circuit with $\hat H_{C'}$ generally does not apply to the problem of a different $\hat H_{C}$. Intuitively, the QAOA quantum circuit performs as an interferometer, where only $\bm s$ at the minima of $\hat H_{C}$ interfere constructively through the $e^{-i\hat H_{C} x_m}$ driving layers \footnote{The saddle points and maximums are ruled out by the cost function.}, especially with a large and smooth neighborhood. We will illustrate related numerical examples later and in Appendices. It may be viable to apply a simpler $\hat H_{C'}$ that shares the same $\boldsymbol{s}_{gs}$ with $\hat H_{C}$; however, this is generally unpractical as the required $\boldsymbol{s}_{gs}$ is unknown beforehand.

Fortunately, for combinatorial optimization problems,
the Hamiltonian $\hat{H}_{C'}=\sum_{i\in C'} \hat H_{i}$ with a partial set $C'\subset C$ offers an answer.  As we mentioned earlier, dropping out clauses improves the energy landscape of a harder problem while ensuring  $\boldsymbol{s}_{gs}$ is the ground state of  $\hat{H}_{C'}$. The caveat is, in addition to $\boldsymbol{s}_{gs}$,
there could be false solutions due to the now fewer constraints. To avoid the false solutions, we substitute these simpler problems $\hat H_{C_{1,\cdots,p}}$ into Eq.~\ref{eq:qaoa_appr} while keeping the cost function in Eq. \ref{eq:qaoa_cost}.

Let us summarize our improvements to the regular QAOA via quantum dropout (Fig. \ref{fig:qaoa}). We start the problem with an efficient classical solver.
If the result is satisfactory, we stop the procedure since there is no point in a quantum solver. Otherwise, these failed classical results,
typically low-lying excited states (local minima), offer insights as we prepare quantum dropout for QAOA: whether a clause should be kept or available for quantum dropout to underweight the distracting local minima and enhance the chances to locate $\boldsymbol{s}_{gs}$. Finally, we optimize $\ket{\mixParam, \drvParam}$ with respect to the original cost function $\langle \hat{H}_C\rangle$ with a complete set of clauses to ensure the uniqueness of the global minimum.
The current procedure does not incur obvious overhead to the conventional QAOA since the preliminary approaches and the quantum-dropout controls
are both inexpensive on a classical computer; see Appendix D. We emphasize that there are essential differences between quantum dropout and dropout in artificial neural networks, and their similarity is merely symbolic: there are no neurons in QAOA's quantum circuit, and quantum dropout operates on the Hamiltonian level; also, we apply quantum dropout at the beginning - remove clauses from the Hamiltonian for a modified QAOA circuit model - and keep the architecture in training and application afterward. These differ from classical dropout, which randomly sets aside a fraction of the neurons during training \cite{MLbook, Nitish2014}.

\begin{figure}
\includegraphics[width=.80\linewidth]{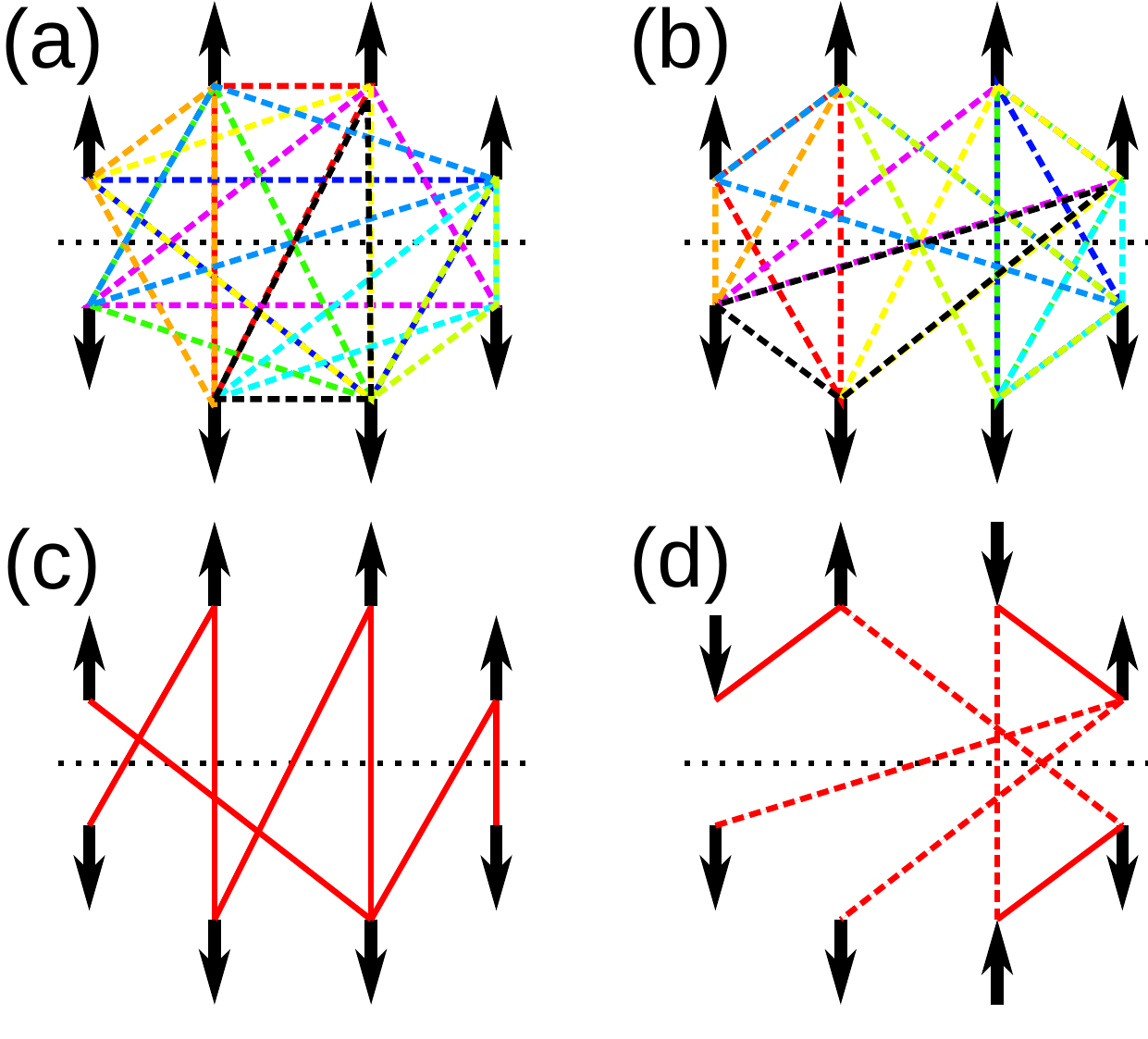}
\caption{(color online)(a) and (b): The NAE3SAT problem with 10 clauses over $n=8$ spins is simpler (harder) when the clauses are more sporadic
(concentrated on a few pairs). Each color denotes a different clause, which must traverse the central dotted line for the not-all-equal condition.
(c) and (d): a step-by-step local analysis first follows the strongest antiferromagnetic interactions (red lines), then the second strongest (red dashed lines),
and so forth. While the result for the simpler problem is consistent with $\boldsymbol{s}_{gs}$, the result for the harder problem largely misleads.}\label{fig:3sat}
\end{figure}

\emph{Not-all-equal 3-SAT problems}\textemdash We use the NAE3SAT problems to demonstrate how to implement QAOA with quantum dropout and its effectiveness due to the straightforward control of their hardness following an intuitive, simple picture, as we explain in the following. This choice will not cause a loss of generality as the NAE3SAT problems are NP-complete, i.e., any quadratic unconstrained binary optimization problem of the decision version \footnote{The decision version asks whether the objective function's minimum could be lower than a given value.} can be reduced to it, yet a polynomial algorithm is not available in general \cite{Moret1988, Dinur2005,garey1979computers, NPcomplete1990}; see Appendix J for further details and examples.

The solution $\boldsymbol{s}_{gs}$ of an NAE3SAT problem satisfies not-all-equal $s_i, s_j, s_k$ for a given set of clauses $[i, j, k]\in C$.
For instance, clause $[1, 2, 3]$ allows $s_1=s_2=1, s_3=-1$ but not $s_1=s_2=s_3=1$. Therefore, we may regard the solution $\boldsymbol{s}_{gs}$ as the ground state of the following Hamiltonian:
\begin{eqnarray}
\hat H_C &=& \sum_{[i,j,k] \in C} \left[ \left( s_i+s_j+s_k \right)^{2}-1 \right]/2 \nonumber \\
&=& \sum_{[i,j,k] \in C} \left(s_is_j + s_is_k + s_js_k \right) + \mbox{const.}\,,
\label{eq:Ham}
\end{eqnarray}
where the interaction between a pair of Ising spins $s_is_j$ is antiferromagnetic, and its strength depends on the number of times $i$ and $j$ appear in pairs within all clauses.
A clause favors antiferromagnet as it imposes two opposite and only one parallel alignment. Therefore, a straightforward and physically intuitive solution is to anti-align the pair of spins with the most appearances in clauses, then the pair with the second most appearances, and so forth, in analogy with the greedy algorithm (Fig. \ref{fig:3sat}c) \cite{Jungnickel1999}. However, if such a local perspective yields globally inconsistent deductions with $\boldsymbol{s}_{gs}$, e.g., multiple pairs of spins with repeated appearances in clauses are counter-intuitively aligned,
see Figs. \ref{fig:3sat}b and \ref{fig:3sat}d, the NAE3SAT problem is commonly harder. We emphasize that despite their restrictive guidelines and thus overshadowed percentage by random problems (Fig. \ref{fig:3sat}a), these challenging problems determine the categorical complexity and are more meaningful from a quantum-solver perspective.

Following these guidelines in Fig. \ref{fig:3sat} for simpler or harder problems, respectively, we generate NAE3SAT problems starting from $\boldsymbol{s}_{gs}$ and accumulating consistent clauses $\hat{H}_{i}$ until the ground state of $\hat{H}_C$ is unique (other than a global $s_r\rightarrow -s_r$ symmetry); see Appendix B for details. To quantify each problem's difficulty, we evaluate the chance of finding $\boldsymbol{s}_{gs}$ in Monte Carlo simulated annealing (SA) \cite{SA1993}. For example, we gradually lower the equilibrium temperature from 64, which is sufficiently higher than most barriers, to 0.64, which is significantly lower than elementary excitations, in $O(10^4)$ single-spin Monte Carlo steps for $N=16$ systems, and obtain a success probability $>95\%$ for most problems following Fig. \ref{fig:3sat}a and $<10\%$ for selected problems following Fig. \ref{fig:3sat}b \footnote{While most harder problems see a much lower $R$ than the simpler ones, we cherry-pick those hardest ones with $R<10\%$ for further analysis.}. The qualitative nature of the energy landscapes of such problems, as well as the effects of quantum dropout, have been illustrated previously in Fig. \ref{fig:landscape}. Next, we examine our numerical results on QAOA on these NAE3SAT problems.

\begin{figure}%[t]
    \begin{subfigure}{.85\linewidth}
        \centering
        \includegraphics[width=\linewidth]{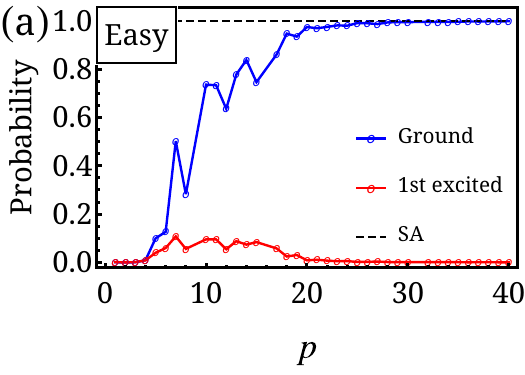}
    \end{subfigure}
    \vfill
  \begin{subfigure}{.85\linewidth}
        \centering
        \includegraphics[width=\linewidth]{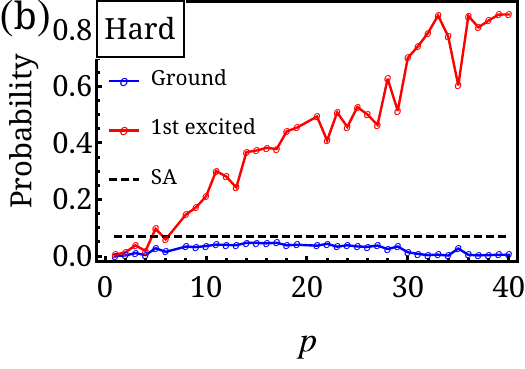}
    \end{subfigure}
    \caption{(color online) We use the probability of achieving the ground state $\boldsymbol{s}_{gs}$ as a measure of the performance of QAOA and SA on an (a) simpler and (b) harder NAE3SAT problem of system size $n=16$. While QAOA performs satisfactorily on the simpler problem $\hat{H}^{E}_{C}$, especially for sufficient circuit depth $p$, it faces challenges on the harder problem $\hat{H}^{H}_{C}$ and performs no better than SA. We estimate the success probabilities over 1000 trials for SA and after 500 steps for QAOA.
    }\label{fig:n16_qaoa}
\end{figure}

\emph{Results} \textemdash First, we apply regular QAOA on typical simpler and harder NAE3SAT problems, whose results are summarized in Fig. \ref{fig:n16_qaoa}. Our QAOA employs the LBFGS algorithm with a learning rate of 0.01. Indeed, QAOA is successful on a simpler problem $\hat{H}^{E}_{C}$, with the ground state's weight tending to $100\%$ as the circuit depth increases; however, such success is less exciting as SA also achieves $\boldsymbol{s}_{gs}$ with a high probability of $\sim 100\%$. On the other hand, QAOA performs no better than SA on a harder problem $\hat{H}^{H}_{C}$, and to make the matter worse, a deeper circuit hardly improves its performance and may even become harmful, as more and more weights get stuck in low-lying excited states. Further, our initializations following heuristics with a linear adiabatic schedule do not help with the difficulties. These behaviors are general to other simpler and harder problems as well; see Appendix A.

\begin{figure}%[t]
    \centering
    \includegraphics[width=.9\linewidth]{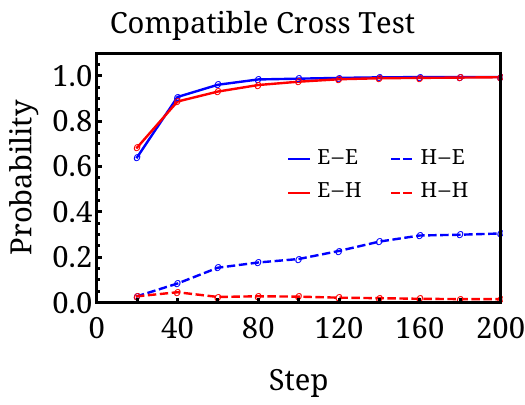}
    \caption{(color online) We evaluate the QAOA performance with $p=30$ via the probability of achieving the ground state $\boldsymbol{s}_{gs}$ as we graft the quantum circuit of $\hat{H}_{C'}$ to the cost function of $\hat{H}_C$, where $\hat{H}_{C'}$ and $\hat{H}_C$ can each be a simpler problem $\hat{H}^{E}_{C}$ or a harder problem $\hat{H}^{H}_{C}$ with the same $\boldsymbol{s}_{gs}$. The horizontal axis labels the number of optimization steps. The legend $X-Y$ denotes $\hat{H}^{X}_{C}$ for the quantum circuit and $\hat{H}^{Y}_{C}$ for the cost function, where $X, Y = E, H$.
    }\label{fig:crossloss}
\end{figure}

As a given problem enters both the quantum circuit and the cost function of QAOA, to locate the difficulty, we perform a cross test where the problem $\hat{H}_{C'} = \hat{H}^E_{C}, \hat{H}^H_{C}$ used in the quantum circuit may differ from $\hat{H}_{C} = \hat{H}^E_{C}, \hat{H}^H_{C}$ in the cost function. We note that $ \hat{H}^E_{C}, \hat{H}^H_{C}$, as studied in Fig. \ref{fig:n16_qaoa}, are relatively simpler and harder yet possess a consistent $\bm{s}_{gs}$ - QAOA generally fails when these two problems are incompatible, see Appendix C. We summarize the convergence of $\ket{\mixParam, \drvParam}$ towards $\ket{\boldsymbol{s}_{gs}}$ in Fig. \ref{fig:crossloss} as a measure of difficulty QAOA faces. The QAOA performs well as long as the quantum circuit engages a simpler problem $\hat{H}_{C'} = \hat{H}^E_C$, and vice versa, while the cost function $\hat{H}_{C}$ plays a relatively minor role, which demonstrates that the quantum circuits is the bottleneck and should be our main target of simplification.

As discussed earlier, quantum dropout rightfully addresses such concerns on QAOA quantum circuits, providing us with simpler yet still compatible $\hat{H}_{C'}$ for the driving layers. As we checked the difficulty of the harder problem $\hat{H}^H_C$ via SA in Fig. \ref{fig:n16_qaoa}, we have also obtained, as a by-product, $29$ low-lying excited states, which help us to choose the dropout clauses more selectively. For example, we only implement quantum dropout on clauses that observe no violation with these distracting local minima; see Appendices D and E. The resulting energy landscape was previously illustrated in Fig. \ref{fig:landscape}d. We leave the cost function $\hat{H}_C=\hat{H}^H_C$ intact with all of the clauses and perform the QAOA optimization following the procedure in Fig. \ref{fig:qaoa}.

\begin{figure}%[t]
    \begin{subfigure}{.85\linewidth}
        \centering
        \includegraphics[width=\linewidth]{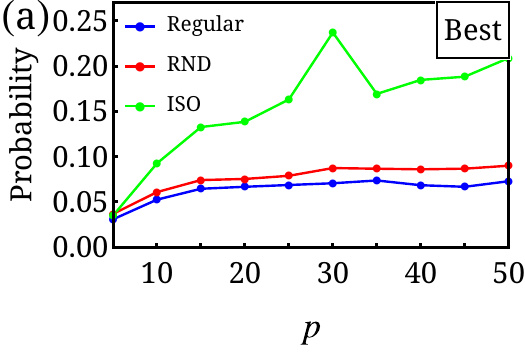}
    \end{subfigure}
    \vfill
    \begin{subfigure}{.85\linewidth}
        \centering
        \includegraphics[width=\linewidth]{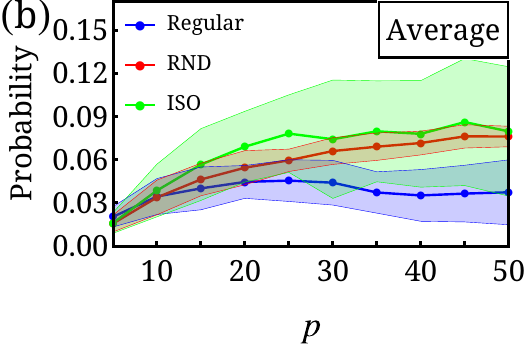}
    \end{subfigure}
    \caption{(color online) By the success probability versus the circuit depth $p$ of (a) the best case and (b) the averages and standard deviations (as the shaded ranges) over the trials, we compare the performance of QAOA of three forms: the regular QAOA (blue line), the QAOA with a quantum dropout of a uniform $\hat H_{C'}$ (green line) or different $\hat H_{C_{1, \cdots, p}}$ (red line) over the driving layers. For each trial with quantum dropout, we randomize the initial parameters $(\mixParam, \drvParam)$ among $(-\pi, \pi)^{\otimes 2p}$ and the clauses for the dropout Hamiltonians, and evaluate the success probability after 200 optimization steps. The number of trials varies according to $p$, $p\leq 20$: $100$, $25\leq p\leq 40$: $50$, and $45\leq p$: $30$.
    }\label{fig:n16_hard_drop}
\end{figure}

First, we set a uniform quantum dropout, randomly ditching $R=50\%$ of the clauses in the available subset, to all driving layers, i.e., $\hat H_{C_1} = \hat H_{C_2} = \cdots = \hat H_{C_p}$. The QAOA performance is shown as the green lines in Fig. \ref{fig:n16_hard_drop}. We observe an evident improvement in favor of quantum dropouts, which becomes more significant as the circuit depth $p$ grows. At $p=50$, the QAOA's probability of locating the ground state doubles on average with the implementation of quantum dropouts, with the best-case scenario offering a success probability of $\sim 0.21$, well exceeding that of $\sim 0.073$ for the regular QAOA and the SA probability of $\sim 0.069$. We do not claim to establish the quantum advantage, as we can employ similar dropout ideas in classical SA to lift its performance; see Appendix G. Intuitively, when $p$ is small, the limiting factor is the quantum circuit's capacity; as $p$ and thus the quantum circuit's expressibility increases, the bottleneck switches to the optimization of the variational parameters, where the regular QAOA commonly gets bogged down and quantum dropout begins to shine (see the averaged performance in Fig. \ref{fig:n16_hard_drop}b).

We also examine the scheme of setting driving layers with different dropouts. The corresponding result is summarized in red in Fig. \ref{fig:n16_hard_drop}, indicating improvements over the regular QAOA at sufficiently large $p$, especially with more aggressive dropout ratios $R$; see Appendix F. We also observe a lower performance variance than the uniform quantum dropout, more subjected to the random dropout configurations; therefore, this dropout architecture is advisable if the number of trails is rather limited.

%If projections were available, combinatorial optimization problems would become straightforward to solve. However, a general classical algorithm is lacking to carry out projections, hampered by the enormous candidate search space.
Combinatorial optimization problems are thought hard to solve efficiently. With simulated annealing, one seeks the global minimum through random exploration and energy comparison.
Interestingly, essentially a quantum interferometer, the QAOA circuits with different dropouts over driving layers may work through a focusing effect on $\boldsymbol{s}_{gs}$: different clause sets lead to different energy landscapes and minima, whose configurations receive constructive interference and enhanced amplitudes. Being the only common minimum of all $\hat{H}_{C'}$ irrespective of the dropouts, $\boldsymbol{s}_{gs}$ remains stand-out through all driving layers. Unfortunately, limited by the current system size ($n=16$) and circuit depth ($p=50$), we have yet to observe an apparent advantage on average for this dropout architecture over the uniform dropout, which requires further studies in future.

\emph{Discussions}\textemdash We have illustrated that while the regular QAOA performs satisfactorily on simpler problems, it still faces significant challenges on meaningful, harder problems. The benefit of a straight increase of the circuit depth quickly saturates, with most of the weights trapped in low-lying excited states. Correspondingly, we have proposed a quantum-dropout strategy for QAOA on harder combinatorial optimization problems, which keeps a number of clauses out of their role in the quantum circuits, therefore easing the landscape of the problem and thus the parameter optimization. The strategy provides an edge over the regular QAOA and SA, especially for harder problems and deeper circuits. For best performance, multiple (quantum dropout) setups and (parameter) initializations should be attempted. Our study also provides valuable insight into the quantum-interfering mechanism of QAOA, which also explains why the model compatibility and simplicity of the quantum circuit are crucial to performance.

Finally, the physical picture for problems' simpler-harder dichotomy and QAOA with quantum dropout straightforwardly apply towards quantum combinatorial optimizations, which may lack a general, compatible classical solver, making the efficient QAOA aided by quantum dropout very useful; we also consider preliminary generalizations to unsatisfiable instances; see Appendices H and I.

\emph{Acknowledgement:} We thank Jia-Bao Wang for insightful discussions. The calculations of this work is supported by HPC facilities at Peking University. YZ and PZ are supported by the National Key R\&D Program of China (No.~2021YFA1401900) and the National Natural Science Foundation of China (No.~12174008 \& No.92270102). ZW and BW are supported by the National Key R\&D Program of China (No.~2017YFA0303302, No.~2018YFA0305602), National Natural Science Foundation of China (No. 11921005), and Shanghai Municipal Science and Technology Major Project (No.2019SHZDZX01).

\section{Appendix A: Additional Examples of Regular QAOA Results on Harder NAE3SAT Problems}

\begin{figure}[t]
    \includegraphics[width=0.95\linewidth]{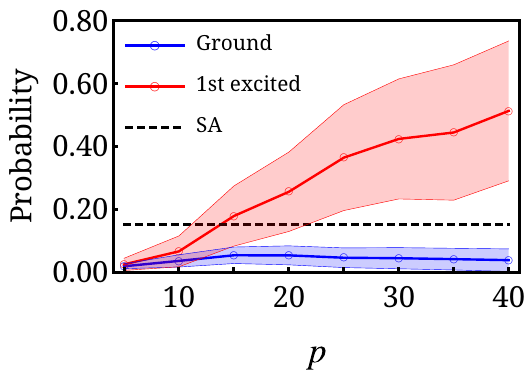}
    \caption{Similar to Fig. 4b in the main text, regular QAOA performs poorly on additional harder NAE3SAT problems generated following the guidelines in the main text. The averages and standard deviations of the outcomes' occupation on the ground states (blue line) and the first excited states (red line) are based upon the $50$ hardest problems pinpointed by SA, with an average success probability displayed as the black dashed line. }
    \label{fig:moreh}
\end{figure}

We illustrate the performance of regular QAOA on $50$ additional harder NEA3SAT problems with system size $n=16$ in Fig. \ref{fig:moreh}. We select these problems according to their SA performances. The results are consistent with and show the generality of our arguments in the main text of Fig. 4b in the main text. Especially, the QAOA performance, the occupation of the target ground state $\bm{s}_{gs}$, stays low and even decreases further at larger circuit depth $p$ to some extent.

\section{Appendix B: Details on NAE3SAT problem generation}

A not-all-equal 3-SAT (NAE3SAT) problem aims to determine the assignment for a set of Boolean variables $\{x_1, \cdots, x_n\}$, $x_r=0, 1$ given a set of clauses $\clauses = \{[i, j, k]: 1\leq i, j, k \leq n, i\neq j \neq k\}$, so that for each of the $M$ clauses, the three variables $x_i, x_j, x_k$ are not all equal, i.e., $x_i + x_j + x_k \neq 0, 3$. This problem is equivalent to the determination of the ground state of a spin Hamiltonian:
\begin{equation}
    \begin{aligned}
    \hat H_C &=\sum_{(i, j, k)\in \clauses} \Big[(\hat s_i + \hat s_j + \hat s_k)^2 - 1\Big]/2 \\
    &=\sum_{(i, j, k) \in \clauses} (\hat s_i \hat s_j + \hat s_j \hat s_k + \hat s_k \hat s_i) + \textrm{const.},
    \end{aligned}
\end{equation}
where the spin operators $\hat s_i$ are the corresponding $\sigma^z$ operators with simple algebraic connection to the qubit $Z$-gates.

Generally, NAE3SAT problems are NP-complete. We can generate such problems randomly and straightforwardly as follows:
\begin{enumerate}
    \item To start with, we choose a configuration as the solution of the problem. In the main text, we use $\ket{\bm{s}_{gs}} = \ket{1}^{\otimes n/2}\otimes \ket{-1}^{\otimes n/2}$. Note that the NAE3SAT problem is symmetric under the all-spin flip $\hat s_r \rightarrow -\hat s_r$.
    \item We randomly generate mutually different $1\leq i, j, k \leq n$. We add the clause $[i, j, k]$ into the set $C$ if it is consistent with the solution.
    \item We repeat step 2 until the number of clauses is sufficient and the solution is unique, i.e., no solution other than $\ket{\bm{s}_{gs}}$ upto the all-spin flip symmetry.
\end{enumerate}
However, as we perform numerical experiments with simulation annealing (SA), it turns out that most NAE3SAT problems with randomly generated clauses are quite simple, even from such a classical algorithm point of view. In SA, we start with a random initial state and a high equilibrium temperature and gradually lower the temperature until it is significantly less than the typical excitation energies, e.g., final temperature $0.5$ over $\sim 20000$ steps for excitation energy $8$ and $n=24$ spins in our case, while keeping Monte Carlo sampling of the spin configurations. By ``simpler'', we mean that the probability of SA locating the target ground state is significantly larger than those harder problems, which have to be generated following a more designated guideline that we summarize as follows:
\begin{enumerate}
    \item To start with, we choose a configuration as the solution of the problem. In the main text, we use $\ket{\bm{s}_{gs}} = \ket{1}^{\otimes n/2}\otimes \ket{-1}^{\otimes n/2}$.
    \item We randomly pick pairs $1\leq i, j \leq n$ whose spins are identical: $s_i = s_j$. These pairs in total should be sub-$n$ that cover at least a portion of the system (beyond measure 0).
    \item For each pair, we add multiple clauses $[i, j, k]$ into the set $C$ if the randomly chosen $k$ has an opposite spin $s_k = -s_i$ and is thus consistent with the solution. The number of clauses per pair should also be sub-$n$ and larger than the average clause number on bonds.
    \item We generate mutually different $1\leq i, j, k \leq n$, especially those dangling sites with no clauses, if any, add the clause $[i, j, k]$ into the set $C$ if it is consistent with the solution, and repeat until the solution is unique, i.e., no solution other than $\ket{\bm{s}_{gs}}$ upto the all-spin flip symmetry.
\end{enumerate}
Note that the non-degenerate requirement is primarily for simplicity and not physically essential.

Behind these guidelines, it is physical intuition that offers a perspective on why such problems are simpler or harder. For the harder problems, the pairs in step 2 received many clauses in step 3 and thus large antiferromagnetic interactions on their bonds, say, $ij$. The SA is a classical algorithm based on the local configuration update and explores the configuration space via local probabilities in the form of a detailed balance between acceptance and rejection. Therefore, SA tends to assign different spins to $i$ and $j$, which opposes the solution $\bm{s}_{gs}$ where $s_i=s_j$. When this happens to a non-negligible number of pairs, the SA becomes misled and trapped in configurations in low-energy neighborhoods that are globally different from $\bm{s}_{gs}$. On the contrary, a randomly generated problem's opposite spins are more likely to accumulate clauses and thus antiferromagnetic interactions than parallel spins following statistics, and such consistent local clues give rise to simpler problems. For reference, SA can reach $\sim 100\%$ accuracy in the simpler problems, while a good portion of the harder problems see $<10\%$ accuracy, indicating the usefulness of physical intuitions.

\section{Appendix C: Failure of QAOA with incompatible circuits}

In the main text, we mention that $\hat H_{C'}$ in the driven layers of the QAOA quantum circuit needs to be compatible with the target problem $\hat H_C$ that we keep in the cost function. Here, we demonstrate the failure of QAOA when $\hat H_{C'}$ and $\hat H_C$ are incompatible.

Let's use the simpler and harder NAE3SAT problems $\hat{H}^E_C$ and $\hat{H}^H_C$ in the main text as our starting point and introduce incompatibility via permutations on the qubits. For example, we consider a permutation $\hat P$ that maps the first qubit to the second, the second qubit to the third, and so forth (the last qubit to the first), so that the solution of $\hat{H}^H_C$ and $\hat{H}^E_C$, $|\boldsymbol{s}_{gs}\rangle=\ket{1}^{\otimes 8}\ket{-1}^{\otimes 8}$ is mapped to $\hat P|\boldsymbol{s}_{gs}\rangle=\ket{-1} \otimes \ket{1}^{\otimes 8} \otimes \ket{-1}^{\otimes 7}$, the solution of $\hat P \hat H^{E,H}_C \hat P^\dagger$. By keeping $\hat H_{C}= \hat{H}^E_C, \hat{H}^H_C$ in the cost function and applying $\hat H_{C'}= \hat P \hat{H}^E_C \hat P^\dagger, \hat P\hat{H}^H_C \hat P^\dagger$ to the driving layers, we introduce incompatibility (a Hamming distance of $2$) between the QAOA circuit and the target problem, which no longer have a common ground state as in the main text.

\begin{figure}[t]
    \includegraphics[width=0.95\linewidth]{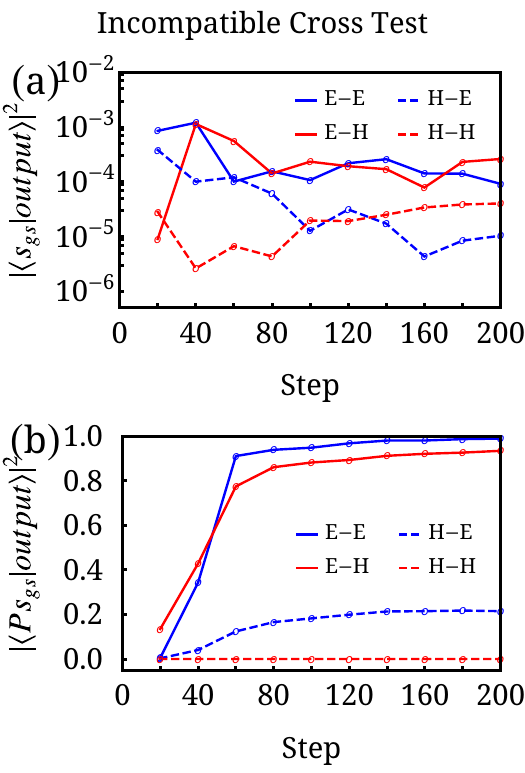}
    \caption{We evaluate the QAOA performance with $p=30$ via the probability of achieving (a) the target ground state $|\bm{s}_{gs}\rangle$ inconsistent with the quantum circuit and (b) the permuted state $\hat P|\bm{s}_{gs}\rangle$ consistent with the quantum circuit (but not the cost function), as we graft an incompatible quantum circuit of $\hat{H}_{C'}$ to the target problem $\hat{H}_{C}$. Note the logarithmic scaling in (a). The horizontal axis labels the number of optimization steps. The legend $X-Y$ denotes $\hat{H}^X_{C'}=\hat{P}\hat{H}^X_C \hat{P}^\dagger$ for the quantum circuit and $\hat{H}^Y_C$ for the cost function, where $X,Y=E,H$.}
    \label{fig:supp_incomp}
\end{figure}

Our main results on the incompatibility tests are shown in Fig. \ref{fig:supp_incomp}, in analogy to Fig. 5 in the main text. The legend $X-Y$ denotes $\hat{P}\hat{H}^X_C \hat{P}^\dagger$ for the quantum circuit and $\hat{H}^Y_C$ for the cost function, where $X,Y=E,H$. Irrespective of such settings, QAOA largely fails to achieve the target ground state $|\boldsymbol{s}_{gs}\rangle$, indicating the necessity of the compatibility of the QAOA circuit.On the contrary, the probability of achieving a non-targeted state is higher - the permuted state $\hat P\ket{\bm{s}_{gs}}$ compatible with the circuit $\hat H_{C'}$, especially when the simpler problem $\hat H^E_{C'}= \hat P \hat{H}^E_C \hat P^\dagger$ is applied. The target cost function $\hat{H}_C$'s incompatibility with $\hat P\ket{\bm{s}_{gs}}$ only costs a minor toll from $\sim 100\%$ to $\sim 90\%$. Overall, our incompatibility tests confirm that the bottleneck of QAOA is the model difficulty at the quantum circuit instead of the cost function.

\section{Appendix D: Procedure of (Heuristic) Quantum Dropout}

Here, we elaborate the detailed procedure of (heuristic) quantum dropout, shown in Fig. 2 in the main text:
\begin{enumerate}
    \item Given a combinatorial optimization problem $\hat H_C = \sum_{i\in \clauses} \hat{H}_{i}$, we start with polynomial-time classical algorithms such as the simulated annealing multiple times. If we obtain the solution, we simply end the whole procedure. However, in case a quantum solver is necessary, the classical solver offers us a set of low-lying excited states $\{\bm{s}_{ex}\}$ that are potential competitors to the target ground state $\bm{s}_{gs}$.
    \item We separate the clauses into two subsets according to each clause's number of violations to $\{\bm{s}_{ex}\}$. For the NAE3SAT example in the main text, the clauses seeing at least one violation are denoted as $\clauses_0$ and kept from dropout, while the rest $\clauses \backslash \clauses_0$ are cached for dropout.
    \item We generate the driving layer Hamiltonians as:
    \begin{equation}
        \hat H_{C'} = \sum_{i \in \clauses'} \hat H_{i}, \ \clauses' = \clauses_0 \cup \textrm{Dropout}_R(\clauses \backslash \clauses_0),
    \end{equation}
    where the function $\textrm{Dropout}_R$ is a random subset of its argument controlled by the dropout ratio $R$ (e.g., $50\%$ in the main text): $\textrm{Dropout}_R(x) \subset x$ and $|\textrm{Dropout}_R(x)| = (1-R) |x|$.
    \item Finally, we optimize the QAOA variational state:
    \begin{equation}
        \ket{\mixParam, \drvParam} = e^{-i \hat H_B \mixParamComp_{p}} e^{-i \hat H_{C_p} \drvParamComp_{p}} \cdots e^{-i \hat H_B \mixParamComp_1} e^{-i \hat H_{C_1} \drvParamComp_1} \ket{+}^{\otimes n},
    \end{equation}
    with respect to its cost function. State $\ket{+}^{\otimes n}$ is the ground state of $\hat H_B = -\sum_{r=1}^n \hat \sigma_{r}^x$. Once our convergence threshold is reached, we measure the optimized $\ket{\mixParam, \drvParam}$ on the $\bm{s}$ basis, which has probability for $\bm{s}_{gs}$ depending on the compositions.
    \item We repeat this procedure until $\bm{s}_{gs}$ is obtained.
\end{enumerate}

\begin{figure}%[b]
    \includegraphics[width=0.98\linewidth]{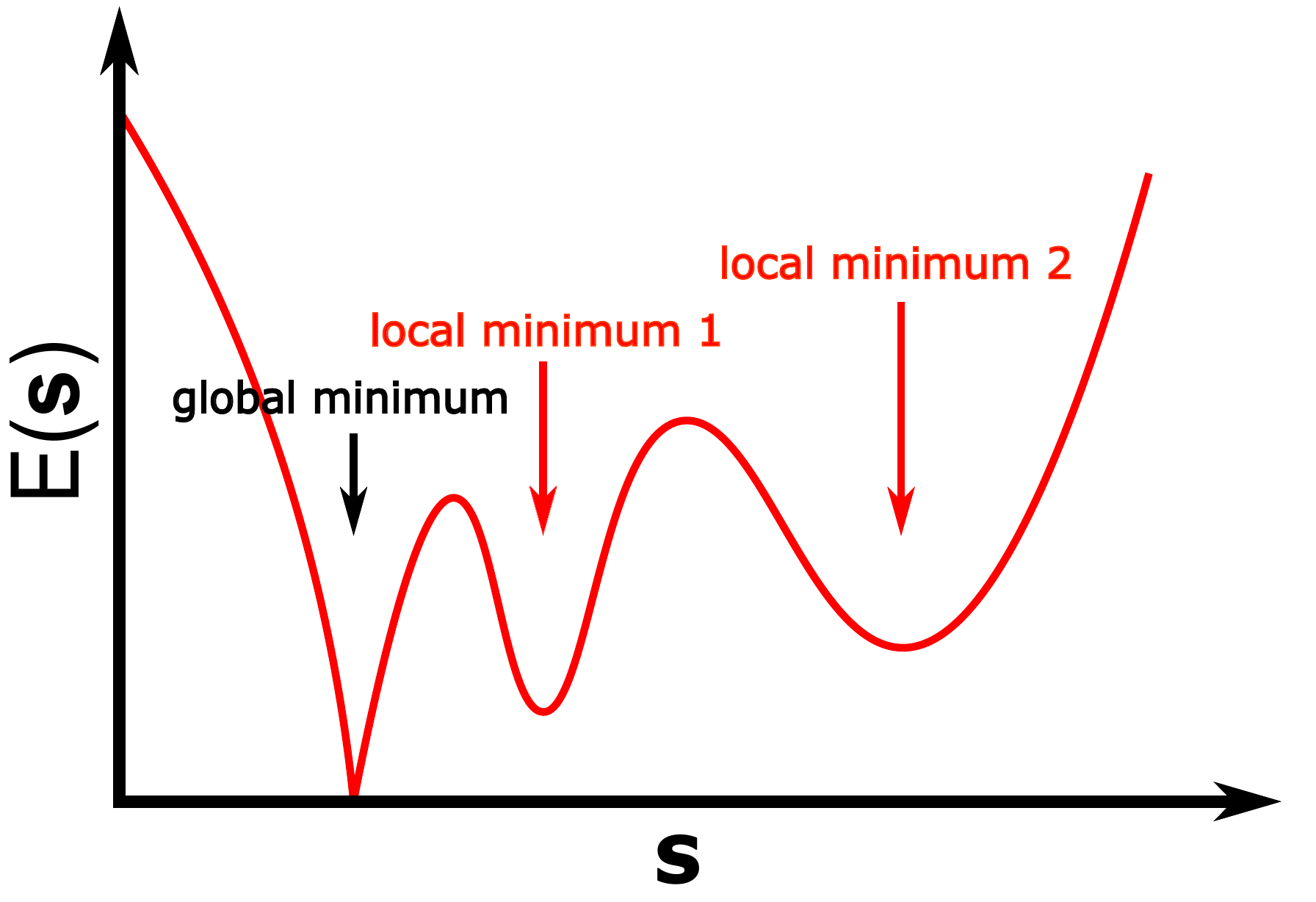}
    \caption{A schematic energy landscape of a harder problem with multiple local minimums. A polynomial classical algorithm such as SA may non-exhaustively identify some of the local minimums, e.g., local minimum 2. The rest of the low-lying excited state, e.g., local minimum 1, if competitive with the target ground state $\bm{s}_{gs}$ and emerge from the results of QAOA before $\bm{s}_{gs}$ is reached, can be included in the heuristics for a step-by-step improvement.}
    \label{fig:harder}
\end{figure}

\begin{figure}
    \centering
    \includegraphics[width=0.98\linewidth]{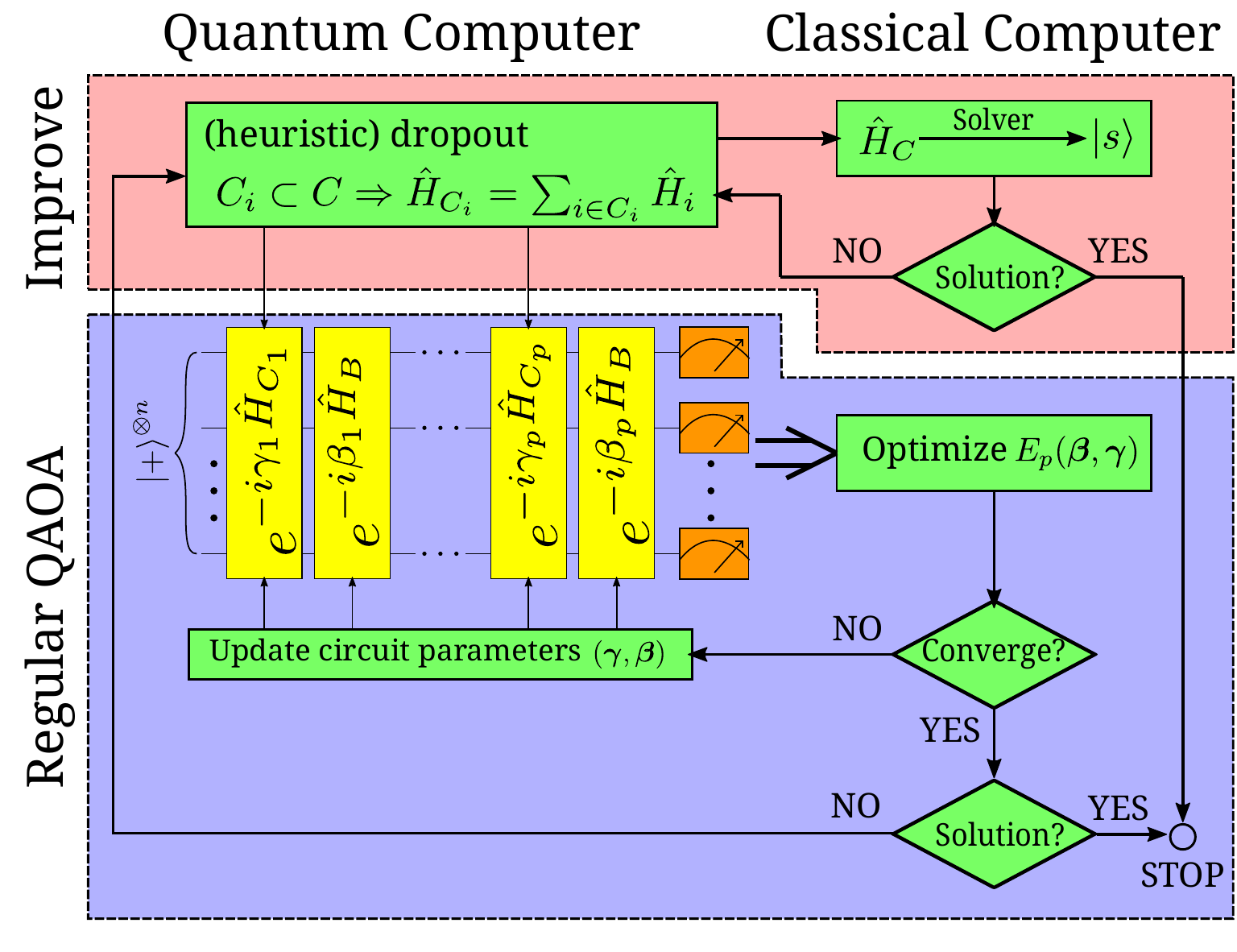}
    \caption{An extended version of the QAOA architecture in Fig. 2 of the main text. If the obtained outcomes from QAOA are not the target state $\bm{s}_{gs}$, we can add these competitive states into $\{\bm{s}_{ex}\}$ for more comprehensive quantum-dropout heuristics for further optimizations.}
    \label{fig:my_label}
\end{figure}

We note that a polynomial-time classical algorithm such as SA may not guarantee an exhaustive set of all competitive local minimums, e.g., Fig. \ref{fig:harder}. On the one hand, we will analyze the impact of heuristics with such a (partial) set in the following sections; on the other hand, if one of these missed-out local minimums emerges from QAOA's measured outcomes before we achieve the target ground state, we can include it into $\{s_{ex}\}$ for improved heuristics, and repeatedly in a step-by-step fashion. We summarize such feedback to quantum-dropout heuristics from unsuccessful QAOA attempts in the extended architecture in Fig. \ref{fig:my_label}.

We build the QAOA circuit and the optimizer with the PyTorch library. The source code and the hyper-parameters for example optimizations will become available upon this letter's publication.

\section{Appendix E: Impact of quantum dropout heuristics}

In the main text (and detailed in the previous section), we introduce a heuristic quantum dropout to simplify the models $\hat H_{C'} = \sum_{i \in C'\subset C} \hat{H}_{i} \ ; \ C' = C_0\cup \textrm{Dropout}_R(C\backslash C_0)$ of the QAOA circuit for the original combinatorial problem $\hat H = \sum_{i\in C} \hat{H}_{i}$, where
\begin{equation}
    C\ni C_0 = \{i: \exists \bm{s}_{ex} \in S \rightarrow H_{i}(\bm{s}_{ex}) = \braket{\bm{s}_{ex}|\hat H_{i}|\bm{s}_{ex}} > 0 \},
\end{equation}
is the subset of all clauses that is violated by at least one low-lying excited state $\bm{s}_{ex}$ obtained via multiple simulated annealing trails. For comparison, we can also perform a random dropout $\hat H_{C'} = \sum_{i \in \textrm{Dropout}_R(C)} \hat{H}_{i}$ without heuristics.

\begin{figure}[t]
    \includegraphics[width=0.95\linewidth]{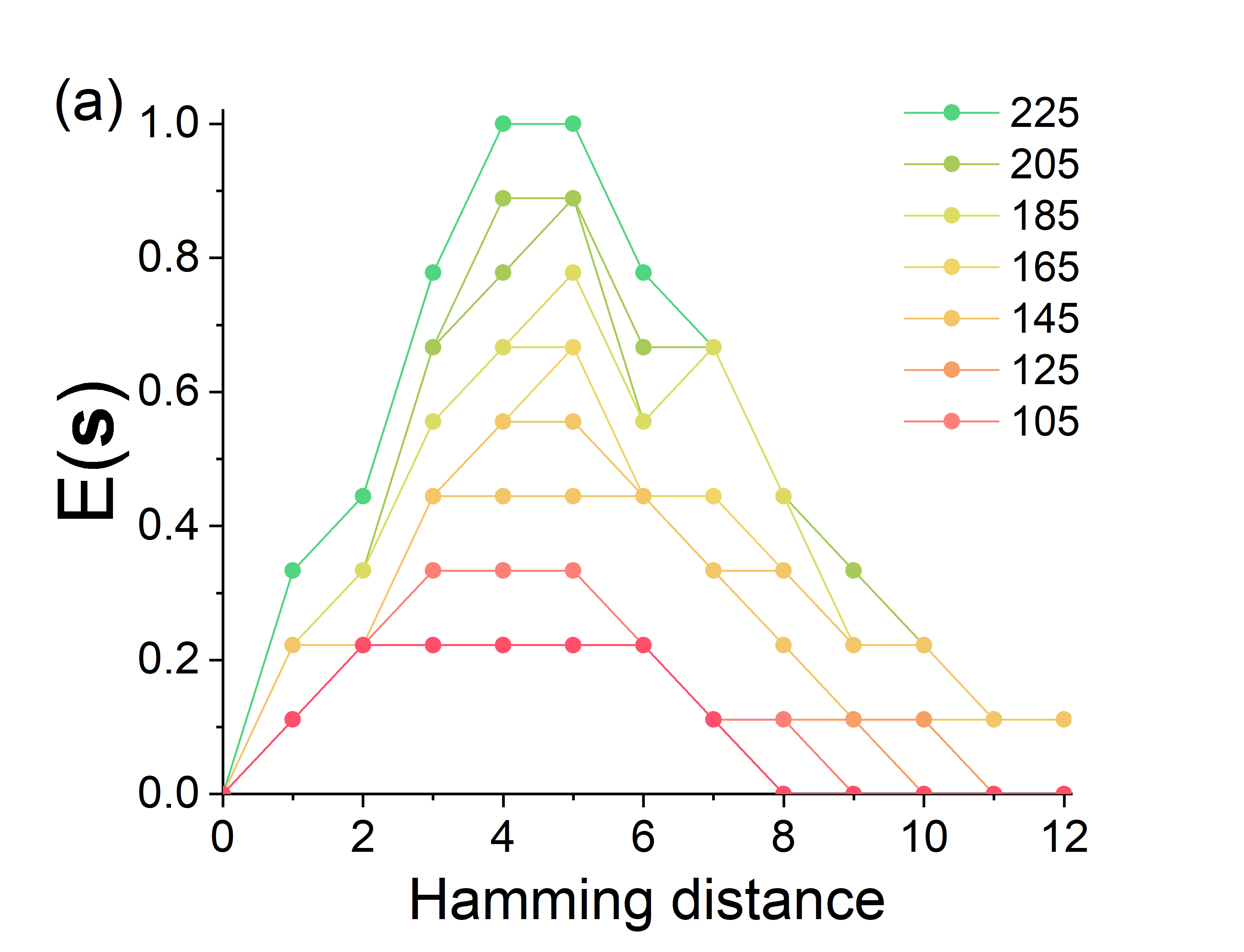}
    \includegraphics[width=0.95\linewidth]{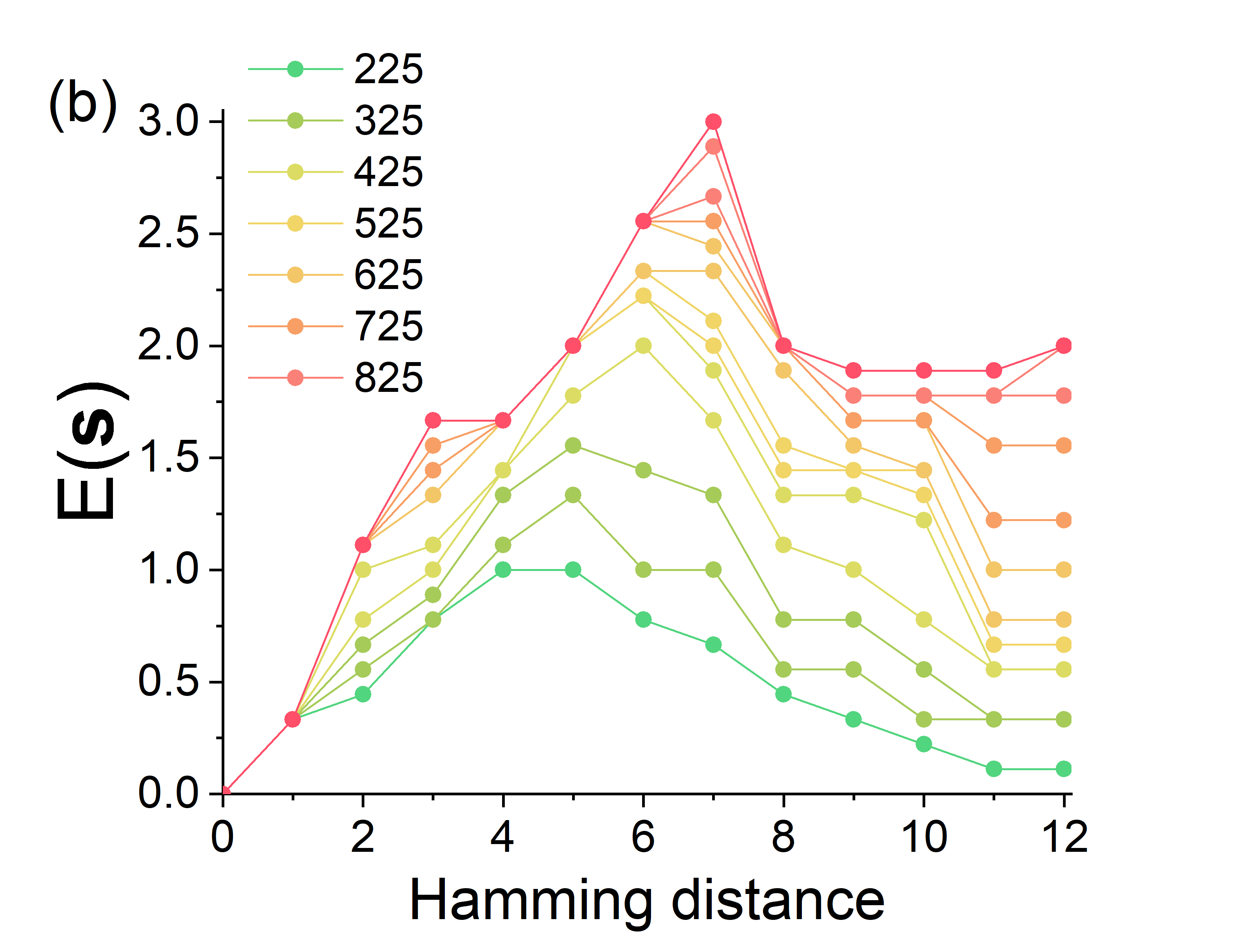}
    \caption{(a) The landscape of a hard problem becomes smoother as more clauses are dropped out, improving the global minimum's standing. In comparison with Fig. 1(d) in the main text, we apply a random quantum dropout without heuristics and observe (many) degenerate zero-energy states even with a smaller dropout percentage. (b) Instead of dropout, we can also improve the energy landscape, to a degree, by adding clauses that violate the competing low-lying excited states. Only the minimum of $E(\bm{s})$ for a specific Hamming distance of $\bm{s}$ from the global minimum is shown for clarity. $n = 24$.}
    \label{fig:supplandscape}
\end{figure}

Quantum dropout generally decreases the energies of the excited states given the fewer remaining constraints while leaving the energy of the target ground state unchanged at zero. It lowers the barriers and thus the difficulties of harder problems; on the other hand, the heuristics help to keep the clauses that distinguish the low-lying excited states, especially those with the most competitiveness, from the target ground state, thus avoiding or at least delaying additional degeneracy and placing a negative bias on these distractions; see Fig. \ref{fig:supplandscape} and Fig. 1(d) in the main text.

\begin{figure}
    \centering
    \includegraphics[width=.95\linewidth]{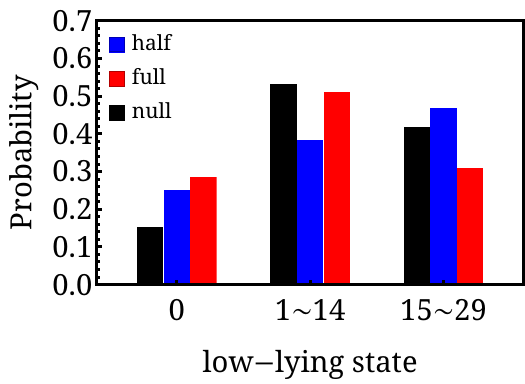}
    \caption{The relative probability of the QAOA outcome on the ground state (denoted as $0$) and the $29$ low-lying excited states located by SA. The data is normalized to show the relative value. We accompany the QAOA with heuristics without heuristics (black bars), with heuristics on half ($1\sim14$) of the low-lying excited states (blue bars), and with all of them as heuristics (red bars). The target problem is the harder NAE3SAT problem $\hat{H}_C^{H}$ we consider in the main text and in the previous sections. The QAOA circuit has the depth of $p=30$.}
    \label{fig:supp_llimp}
\end{figure}

To see how the inclusion of low-lying excited states in quantum-dropout heuristics influences the performance of QAOA, we show the probability of the QAOA outputs over the ground state and the 29 low-lying excited states obtained by SA with different heuristics in Fig. \ref{fig:supp_llimp}: one without heuristics, one with heuristics from half of the low-lying excited states ($1\sim14$), and one with heuristics from all $29$ low-lying excited states. We find that the QAOA outputs on the low-lying excited states included in the quantum-dropout heuristics are suppressed and lowered, with more probability re-distributed to the other low-lying excited states and, importantly, the target ground states - especially when we incorporate more competitive low-lying excited states. In summary, though optional, the heuristics are consistent with our physical intuition and helpful for their negative biases on the low-lying excited states and enhanced competitiveness of the target ground state in QAOA. We will discuss the robustness of the QAOA performance to such variability in the quantum-dropout heuristics in the next section.

We note that quantum dropout is not the only way to ease the landscape. Given the set of competitive low-lying excited states, we can also opt for more weights on clauses disagreeing with them and increase their energies. In practice, however, it is difficult to guarantee an exclusive list of all competitive low-lying excited states, which may be necessary to improve the energy landscape significantly. In Fig. \ref{fig:supplandscape}(b), we show the impact of such an approach on the same NAE3SAT problem as in Fig. \ref{fig:supplandscape} and Fig. 1(d). On the other hand, more clauses usually need a higher-precision platform. Therefore, such an alternative's inefficiency and additional cost make quantum dropout a more suitable implementation in experiments.

Besides, we have studied soft quantum dropout, where we allow each clause to contribute to $\hat{H}_{C'}$ in variable percentages instead of the binary assignment of in or out. In theory, this setup may ravel up the spectrum leading to more destructive and inconsistent interference through the driving layers for the low-lying excited states. However, we have not observed numerical evidence supporting its advantages yet.

\section{Appendix F: Robustness of Heuristic Quantum Dropout to the Number of Distractions}

There are two core hyper-parameters controlling the quantum dropout of the algorithm: the dropout ratio $R$ and the number of competitive low-lying excited states $l$ incorporated in heuristics - violated by a number of clauses, $f(l)$, concave with respect to $l$ statistically: $f(\alpha l) \geq \alpha f(l)$ for $\alpha \in [0, 1]$. As a result, the dropout reduces the problem with $M$ clauses to $(1-R)[M-f(l)] + f(l)$ clauses; see previous sections. In the main text, we showcase QAOA performance examples with a harder NAE3SAT problem over $n=16$ spins and $M = 112$ clauses with a dropout ratio of $R=0.5$ and heuristics from all of the $l=29$ low-lying excited states obtained from SA, ending up with $\sim 70$ clauses.

\begin{figure}[t]
    \includegraphics[width=0.95\linewidth]{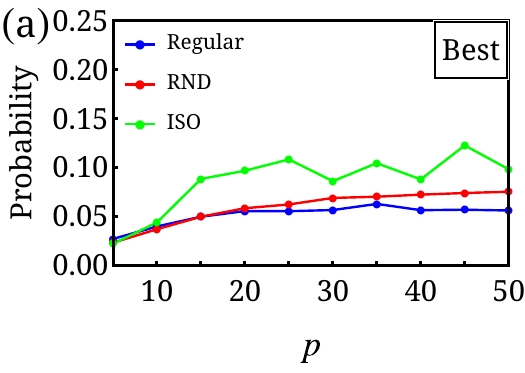}
    \includegraphics[width=0.95\linewidth]{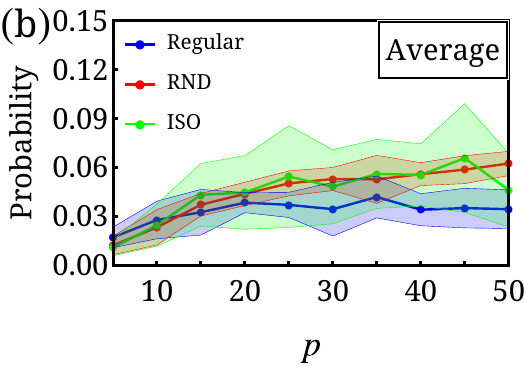}
    \caption{The success probability of (a) the best case and (b) the averages and standard deviations over a number of trials versus the QAOA circuit depth, as we halve the number of low-lying excited states for heuristics, $l=14$, shows the robustness of quantum dropout's advantage towards $l$. The number of trails is also half of those in Fig. 6 in the main text, while the rest of the setup is the same. $R=0.5$.}
    \label{fig:supp_half}
\end{figure}

A natural question is the impact of $l$ on and the robustness of the overall performance, as it is difficult to guarantee an exhaustive search for all competitive low-lying excited states; see Fig. \ref{fig:harder} and the related discussion. For example, for $l=14$ around half of what we use in main text, the number of clauses in $C_0$ is reduced from $f(29) = 22$ to $f(14)\sim 16$. The corresponding results are shown in Fig. \ref{fig:supp_half}. We find the performance of dropout QAOA is quantitatively similar to the result of $l=29$ in the main text, beating both the regular QAOA and SA. Although the average success probabilities of QAOA with a quantum dropout of an identical $\hat H_{C'}$ or different $\hat H_{C_{1, \cdots, p}}$ over the driving layers are $0.076$ and $0.077$, respectively, slightly less than $0.076$ and $0.079$ for $l=29$ in the main text, the difference is small and below the level of uncertainty. Therefore, it suffices to say that our quantum dropout strategy is robust to $l$.

\begin{figure}[t]
    % in the supp, r, the dropout ratio, is the ratio of dropouted clauses. In the filename, it is the preserved ratio.
    \includegraphics[width=0.95\linewidth]{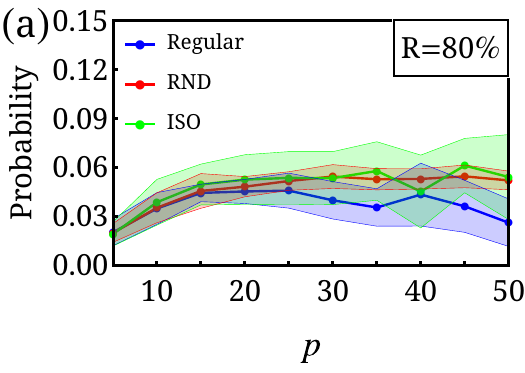}
    \includegraphics[width=0.95\linewidth]{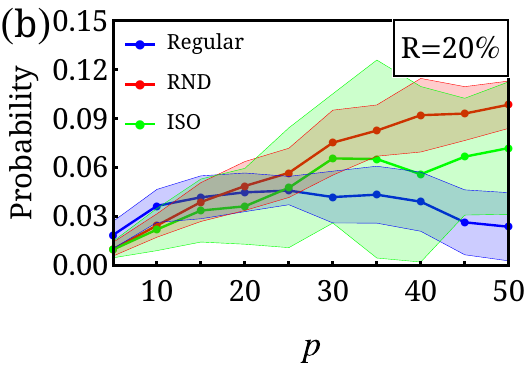}
    \caption{We compare the average success probability and its standard deviation versus the model depth of QAOA with different quantum dropout ratios: (a) $R=0.2$ and (b) $R=0.8$. The rest of the settings are the same as Fig. \ref{fig:supp_half}(b). $l=14$.}
    \label{fig:supp_other_r}
\end{figure}

We also examine how the other hyper-parameter, the dropout ratio $R$, impacts the QAOA performance. We summarize the results for $R=0.2$ and $R=0.8$ in Fig. \ref{fig:supp_other_r}, while the results for $R=0.5$ are in Fig. \ref{fig:supp_half} as well as Fig. 6 in the main text ($l=29$). Intuitively, a larger $R$ keeps more clauses and is closer to the regular QAOA without quantum dropout. It reduces the variations and the extent of improvement on the energy landscape, thus the performance. The results in Fig. \ref{fig:supp_other_r}(a) agree with our intuition, as the performances exhibit a closer resemblance between different QAOA setups. On the other hand, a smaller $R$ dropouts a larger portion of clauses, enhancing the potential benefits of quantum dropout. However, depending on the specific clauses choice, the dropout may introduce degeneracy to the problem and increase diversity. Therefore, more trials may be necessary to utilize the benefits fully. Indeed, the results in Fig. \ref{fig:supp_other_r}(b) show larger average success probabilities. Interestingly, the QAOA with a quantum dropout of different $\hat H_{C_{1, \cdots, p}}$ over the driving layers is more capable of utilizing a lower $R$ ratio and suppressing the variance at the same time.

Finally, we note that the regular QAOA tends to perform worse with increasing circuit depth $p$ beyond larger $p\geq 40$, in Figs. \ref{fig:supp_half}, \ref{fig:supp_other_r}(a) and  \ref{fig:supp_other_r}(b). It is less apparent in Fig. 6(b) in the main text; however, there is some extent of randomness, and we have refrained ourselves from cherry-picking results to support a particular claim. If this tendency is general, deeper QAOA circuits are NOT the solutions for harder problems due to the difficulties commonly associated with the optimizations, which our quantum dropout strategy may largely alleviate. In short, QAOA with quantum dropout is more capable of utilizing a deeper circuit for more meaningful harder problems.

%Add for discussion about classical analog of quantum dropout
\section{Appendix G: Impact of dropout on simulated annealing performance}

As quantum dropout significantly eases the energy landscapes of harder problems, it is understandably capable of boosting the performance of classical simulated annealing as well. We carried out classical SA algorithms on a harder problem ($\sim 7\%$ success rate in regular SA) $H_C$ with an $R=50\%$ (heuristic) dropout of its clauses, and the success rate indeed increased significantly to $7\sim 35\%$ with a large standard deviation. Such performances are similar to QAOA with a quantum dropout of a uniform $H_C$ over the driving layers (green color in Fig. 6b). It is still unclear whether a classical analogy of QAOA with different dropouts over driving layers exists or not.

\begin{figure}[t]
    \centering
    \includegraphics{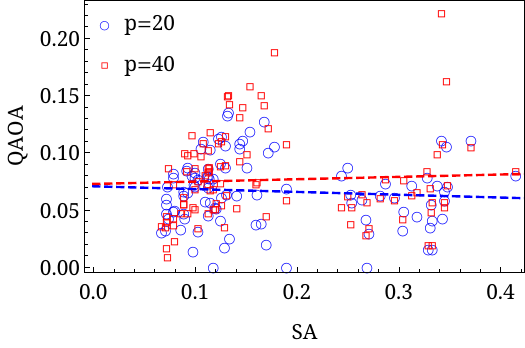}
    \caption{The success probabilities of QAOA (ISO scheme) and classical SA suggest a lack of clear correlation over 100 $H_C$  with quantum dropout. We obtain the SA success probability via 1000 trials, each of which lowers the temperature from 100 to 0.05 in 2000 steps. The QAOA success probability equals the $s_gs$ weight, where we consider circuit depth of $p=20$ (circle) and $p=40$ (square). The dashed lines are linear regressions.}
    \label{fig:supp_dropsa_lr}
\end{figure}

Though quantum dropout also enhances SA performance, we find that the classical (SA) and quantum (QAOA) favor vastly different dropout configurations. For example, we study 100 $H_C$ following the harder problem in the main text (with $N=16$ and $\sim 7\%$ success probability in regular SA) and different (heuristic) dropouts. The performances of SA and QAOA (with uniform dropout) show little correlation; see Fig.~\ref{fig:supp_dropsa_lr} and Table. \ref{tab:supp_corr}, suggesting that quantum and classical algorithms function differently and are not entirely analogous.

\begin{table}[t]
    \centering
    \begin{tabular}{cc}
    \hline
    QAOA depth $p$ & Correlation coefficient \\
    20 & -0.0785 \\
    30 & -0.0082 \\
    40 & 0.0512 \\
    \end{tabular}
    \caption{The Pearson correlations between the successes of classical SA and QAOA with circuit depths $p=20, 30, 40$ on various harder problems $H_C$ with consistent quantum dropout.}
    \label{tab:supp_corr}
\end{table}

\section{Appendix H: Quantum combinatorial optimization problems}

Here, we give an example of quantum combinatorial optimization problems:
\begin{eqnarray}
\hat{H} &=& \sum_{[ijk]\in C} \left(\vec{S}_i+\vec{S}_j+\vec{S}_k\right)^2  /2 \label{eq:qcom}\\
&=& \sum_{[ijk]\in C} \left(\vec{S}_i\cdot\vec{S}_j + \vec{S}_i\cdot\vec{S}_k + \vec{S}_j\cdot\vec{S}_k \right) + const., \nonumber
\end{eqnarray}
whose ground state is a direct product of dimers (singlets) $|gs\rangle = \prod_{[mn]\in D} |\uparrow\rangle_m |\downarrow\rangle_n - |\downarrow\rangle_m |\uparrow\rangle_n$.
In analogy to the Majumdar-Ghosh spin-chain model \cite{MGchain1969}, each clause serves as a projection operator onto the total spin $1/2$ sector of the three included spins; when a spin-singlet exists between either $ij$, $ik$, or $jk$ for a clause, the total spin $\vec{S}_i+\vec{S}_j+\vec{S}_k$ becomes $1/2$, and the clause is satisfied.

Given the set of clauses $C$, the optimization target is to determine $|gs\rangle$ and its dimer configuration $D$ within. In comparison with the NAE3SAT problems in the main text, the Hamiltonian in Eq. \ref{eq:qcom} is quantum due to the Heisenberg interactions, thus $\hat{H}_{i}$ for each clause no longer commutes with each other and a classical solution is generally not available. Still, the target ground state $\ket{gs}$ individually satisfy (the ground-state condition for) each $\hat{H}_{i}$. On the other hand, QAOA \cite{Hsieh2019} as well as our quantum dropout strategy can readily apply to such quantum optimization problems without any modifications or complications.

Similar to the NAE3SAT problems, the quantum combinatorial optimization problems in Eq. \ref{eq:qcom} can be constructed by generating consistent clauses with a predetermined $|gs\rangle$ until the number of clauses is sufficient and the ground state is unique. We can also establish relatively harder problems that contradict local intuitions. For instance, a clause signals the existence of a singlet-pair among its three spins and contributes antiferromagnetic Heisenberg interactions to the three corresponding bonds. Therefore, local intuitions on two overlapping clauses, $[ijk]$ and $[jkl]$, suggest the preference for a singlet over the shared pair $jk$. However, the actual dimer configuration may turn out to be upon pairs $ij$ and $kl$, or $ik$ and $jl$ instead, and if such misleading persists throughout the system, the conclusions may end up globally different from $|gs\rangle$.

\section{Appendix I: Generalization to unsatisfiable problems}

\begin{figure}
    \centering
    \includegraphics[width=.9\linewidth]{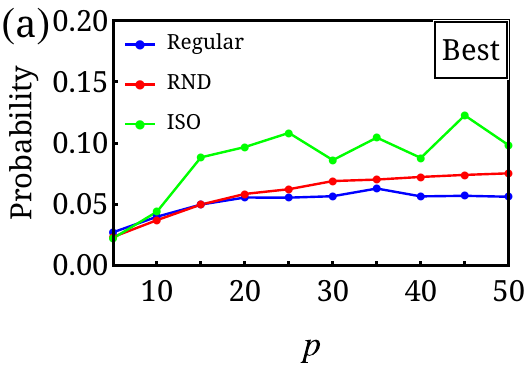}
    \includegraphics[width=.9\linewidth]{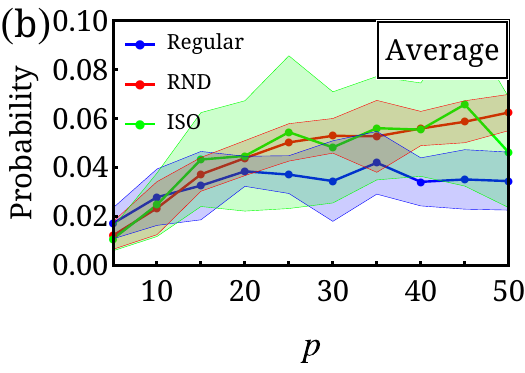}
    \caption{The success probability of (a) the best case and (b) the averages and standard deviations over a number of trials versus the QAOA circuit depth for a NAE3SAT problem that is unsatisfiable . The quantum dropout is based upon the heuristics over 15 excited states obtained by SA. The dropout ratio $R=0.5$ and the number of trials are the same as in Fig. \ref{fig:supp_half}.}
    \label{fig:supp_confl}
\end{figure}

In the main text and the above sections, we focus on the NAE3SAT problems generated with an existing solution, i.e., there is at least one configuration (other than $-\bm{s}_{gs})$ that can satisfy all clauses in a given problem. The corresponding complexity class to locate $\bm{s}_{gs}$ is NP-complete. However, a general NAE3SAT problem may be unsatisfiable. Finding the ground state of the corresponding Hamiltonian in this case is known to be of NP-hard class.

To generate a harder, unsatisfiable NAE3SAT problem, we start from a harder, satisfiable NAE3SAT problem - the primary example of this letter. Then, we append additional clause(s) violating the target ground state and make sure that it has lower energy than all first-excited states, adding further clauses if necessary. The scarcity of the added clauses ensures that the energy landscape is approximately and qualitatively intact.

One of the harder, unsatisfiable NAE3SAT problems we generate has a success probability of $\sim 7\%$ and $31$ low-lying states in SA. However, without a clear knowledge of the satisfiability, we generally have no idea whether these $31$ low-lying states are the target ground state or not. Still, we can efficiently count the number of violations each $\bm{s}$ outcome receives and select the subset of low-lying excited states with violations (energy) above their common minimum - these are definitely the excited states, and we can implement quantum-dropout heuristics based upon them. Here, we obtain $15$ valid excited states out of the $31$ low-lying states.

Another complication due to the unsatisfiability is quantum dropout's threat to circuit compatibility, as the problem $\hat H_{C'}$ after the dropout may have a (globally) different ground state than $\bm{s}_{gs}$, which becomes a low-lying excited state instead since it does not satisfy all clauses in the first place. Still, such a low-lying excited state may form a constructive quantum interference (probably to a lesser degree), and we can also attempt multiple quantum dropout scenarios so that there exist cases in favor of instead of against the target ground state.

We summarize the numerical results on QAOA performance on such a harder, unsatisfiable NAE3SAT problem in Fig. \ref{fig:supp_confl}. A general performance boost from quantum dropout still exists, especially when circuit depth $p$ is sufficient. However, there is a reduced margin over SA, and there are cases where the QAOA success probability becomes worse with quantum dropout. With a quantum dropout of an identical $\hat{H}_{C'}$ over the layers, the success probability reaches $0.123$ at $p=45$, surpassing the regular QAOA at $0.057$ and SA at $0.070$. With a quantum dropout of different $\hat H_{C_{1, \cdots, p}}$ over the layers, QAOA also works well on average, and its lower variance offers an approach more controlled, and useful given the worse lower bound. We also note that heuristics are considerably more helpful on unsatisfiable problems than satisfiable ones.

\section{Appendix J: Equivalence and mapping between problems of equivalent NP complexity class}

In the main text, we have mainly focused on the NAE3SAT problems for demonstrations, given the straightforward control over their hardness. The NAE3SAT problem belongs to the NP-complete complexity class, which implies that if a polynomial-time algorithm exists to solve it, all NP problems can be solved efficiently. This property of NP-completeness provides an equivalence between different forms of NP-complete problems: 3SAT problems, general SAT problems, number partition problems, and graph-covering problems can be mapped to each other with auxiliary qubits and time complexity polynomial to the size of the problem. Such equivalence ensures the universality and generality of NAE3SAT problems, where an algorithm designed to solve NAE3SAT can be utilized to solve other NP-complete problems without intractable costs of resources.

More specifically, we demonstrate the mapping of a Quadratic unconstrained binary optimization (QUBO) problem to an NP-like problem and the mapping of a MaxCut problem to a NAE3SAT problem as follows:

Although QUBO problems are not decision problems with a simple yes or no answer, they are closely related to NP problems. Without loss of generality, for a QUBO objective function $f(x)$ with $x\in \{0,  1\}^{\otimes n}$, we can design a decision problem $D(k, f)$: Is there a bitstring $x$ such that $f(x)\leq k$?

The original QUBO problem becomes a series of $D(k, f)$ with a varying $k$. If the cost of $f(x)$ is polynomial in $n$, we can check whether $f(x)\leq k$ or not given a trial solution $x$, and $D(k, f)$ is at most in the NP class and not "harder" than any NP-complete problem. (If the cost of $f(x)$ is above polynomial in $n$, then the problem is undoubtedly intractable - to the best of our knowledge, any optimization algorithm necessitates the evaluation of the objective function itself.)

Also, we can map a MaxCut problem to a NAE3SAT problem. Let us define a decision problem $C(k; V, E)$ as: Is there a cut of the graph $G(V, E)$ whose size is larger than $k$? Since the answer is undoubtedly No for $k>|E|$, we need to call $C(k; V, E)$ at most $|E|$ times to solve the corresponding MaxCut problem.

Next, we map $C(k; V, E)$ to an SAT problem. Define Boolean variables $v_i$ and $e_{ij}$ on each vertex in $V$ and edges in $E$. A cut of $G$ is thus a subset of $V$ and also an assignment to the language:
\begin{equation}
    \bigwedge_{i, j\in V} \Big(e_{ij} == (v_i \neq v_j)\Big),
\end{equation}
where the operator $==$ determines whether two values are equal, achievable by the XOR operator. The cut is specified by the vertices $i$'s of $v_i=1$. The $e_{ij}$ describes whether edge $(i,j)$ is cut. Thus, the satisfiability version of $C(k; V, E)$ reads:
\begin{equation}
    L = \Big\{\bigwedge_{i, j\in V} \Big(e_{ij} == (v_i \neq v_j)\Big)\Big\} \bigwedge \Big(\sum_{(i, j) \in E} e_{ij} \geq k\Big).
\end{equation}
Finding whether $L$ is satisfiable answers the problem $C(k; V, E)$. The summation of Boolean variables can be implemented by the adder in digital circuits. As a general SAT problem, $L$ cannot be harder than a NAE3SAT problem. Indeed, the existence of such mapping is guaranteed by NP-completeness.

\bibliography{refs}

\end{document}